\documentstyle[aps]{revtex}
\begin{document}
\author{D. Foerster}
\address{CPTMB, URA1537\\
Universit\'{e} de Bordeaux\\
Rue du Solarium,\\
33174 Gradignan, France}
\author{F. Triozon}
\address{Ecole Normale Superieure de Lyon,\\
46 All\'{e}e d'Italie\\
69364 Lyon, France}
\title{The Quantum $O(N)$ Heisenberg Antiferromagnet for $N\rightarrow \infty $ }
\maketitle

\begin{abstract}
We study the $O(N)$ quantum Heisenberg antiferromagnet using a
parametrisation in terms of real fermions. The $N\rightarrow \infty $ limit
of the unfrustrated model is controlled by a saddle point representing
singlets on dimers that cover the lattice and which is infinitely degenerate
in many cases. An infinite degeneracy of the $N=\infty $ ground state occurs
also in some, but not all, frustrated quantum antiferromagnets. Our results
are similar to results reported previously on an $SU(N)$ generalisation of
the Heisenberg antiferromagnet for $N=\infty $.

PACS numbers: 75.10.Jm, 05.30.-d
\end{abstract}

\section*{Introduction}

The idea that the ground state of a frustrated spin 1/2 antiferromagnet
should be describable in terms of products of spin singlets was suggested
originally in the context of triangular spin $\frac{1}{2}$ antiferromagnets 
\cite{PWA}. Products of singlet states are exact ground states of certain
frustrated spin 1/2 antiferromagnetic spin chains with special couplings 
\cite{Majumbdar} and they have been used in variational approaches \cite
{variational} and in heuristic arguments.

In the present note we study the $O(N)$ Heisenberg antiferromagnet in the $%
N\rightarrow \infty $ limit by using a constraint free parametrisation of
the $O(N)$ Heisenberg antiferromagnet in terms of real fermions. We show
that the $N\rightarrow \infty $ limit is dominated by a mean field
configuration with small $\sim \frac{1}{\sqrt{N}}$ fluctuations. We are able
to locate the saddle point of lowest energy in the case of unfrustrated
antiferromagnets and for a subclass of frustrated ones. However, in many
cases we do not find a single saddle point, but instead an infinity of
saddle points that are all degenerate in energy. These saddle points are
essentially singlets on nearest neighbor dimers that cover the lattice and
correspond to the RVB singlets of Anderson.

Our results are similar to those obtained by D.S. Rokhsar on the $%
N\rightarrow \infty $ limit of the $SU(N)$ extension of the Heisenberg
antiferromagnet \cite{Rokhsar}.

\section*{Majorana representation of the O(N) Heisenberg model.}

It is known that operators that operate on spinors can be generalised from
the O(3) group to O(N) by using O(N) Dirac matrices. Somewhat less familiar
is the fact that these Dirac matrices can also be considered to be real
fermions \cite{Tsvelik}, thus providing us with the following representation
of $O(N)$ spin matrices: 
\begin{eqnarray}
s_{k,l} &=&-\frac{i}{2}\left( \eta _{k}\eta _{l}-\eta _{l}\eta _{k}\right) 
\text{, }k,l=1,2,..N \\
\text{ }\left[ \eta _{k,}\eta _{l}\right] _{+} &=&\delta _{kl}\text{, }\eta
_{l}^{+}=\eta _{l}  \nonumber
\end{eqnarray}
The representations of $s_{k,l}$ can be characterised by the value of their
quadratic invariant 
\begin{equation}
s^{2}\equiv \frac{1}{2}\sum_{i,j=1..N}s_{ij}s_{ij}=-\frac{1}{2}\sum_{i\neq
j=1..N}\eta _{i}\eta _{j}\eta _{i}\eta _{j}=\frac{1}{8}\left( N^{2}-N\right) 
\end{equation}
For $N=3$ and $O(3)$ this representation reduces to 
\begin{eqnarray}
s_{i} &=&-i\eta _{k}\eta _{l}\text{, }i,k,l=1,2,3\text{ cyclic} \\
s^{2} &=&\frac{1}{2}(\frac{1}{2}+1)  \nonumber
\end{eqnarray}
a representation that has been discussed extensively in the first of ref\cite
{Tsvelik}. To define the $O(N)$ Heisenberg model, we associate a spin
represented by fermions with each point of a lattice: 
\begin{eqnarray}
s_{x,kl} &=&-\frac{i}{2}\left( \eta _{xk}\eta _{yl}-\eta _{yl}\eta
_{xk}\right)  \\
\left[ \eta _{xk}\text{,}\eta _{yl}\right]  &=&\delta _{kl}\delta _{xy} 
\nonumber
\end{eqnarray}
and use the invariant scalar product 
\begin{equation}
(s_{x}s_{y})_{O(N)}=\frac{1}{2}\sum_{k,l=1}^{N}s_{x,kl}\cdot s_{y,kl})
\end{equation}
to define the interaction:

\begin{equation}
H_{Heisenberg}=\sum_{<x,y>}J_{xy}(s_{x}s_{y})_{O(N)}=\sum_{<x,y>}J_{xy}%
\left[ \frac{N}{8}+\frac{1}{2}\left( \sum_{k=1}^{N}\eta _{x,k}\eta
_{y,k}\right) ^{2}\right] 
\end{equation}
The fermions $\eta _{x,k}$ are real as is appropriate in a situation without
electric charges, and there are no constraints on the physical Hilbert
space. To get familiar with the representation of spins in terms of real
fermions, we may calculate the energy of a pair of nearest neighbor points $%
x,y$ or a ''dimer '' that interact via their spins. We do this most easily
by introducing ordinary complex fermions: 
\begin{eqnarray}
A_{\text{dimer},k} &=&\frac{1}{\sqrt{2}}\left( \eta _{x,k}+i\eta
_{y,k}\right) \text{, }A_{\text{dimer},k}^{+}=\frac{1}{\sqrt{2}}\left( \eta
_{x,k}-i\eta _{y,k}\right) \text{, }k=1,..N \\
\left( s_{x}s_{y}\right) _{O(N)} &=&\frac{N}{8}-\frac{1}{8}\left[
\sum_{k=1}^{N}(A_{\text{dimer,}k}^{+}A_{\text{dimer},k}-A_{\text{dimer,}k}A_{%
\text{dimer},k}^{+})\right] ^{2}  \nonumber
\end{eqnarray}
and find 
\begin{equation}
E_{\text{dimer}}\equiv <\left( s_{x}s_{y}\right) _{O(N)}>=-\frac{N^{2}}{8}+%
\frac{N}{8}  \label{dimer}
\end{equation}
The ground state energy of a dimer is seen to vary smoothly as a function of 
$N$.

\section*{A toy model}

To gain further insight into the nature of the $N\rightarrow \infty $
extrapolation, we calculate, for arbitrary $N$, the ground state energy of
four spins located at the corners of a square with interactions along the
edges and across the diagonals: 
\begin{eqnarray}
H_{square} &=&s_{1}s_{2}+s_{2}s_{3}+s_{3}s_{4}+s_{4}s_{1}+\varepsilon \left(
s_{1}s_{3}+s_{2}s_{4}\right)   \nonumber \\
&=&\frac{1}{2}\left[ \left( s_{1}+s_{3}\right) +\left( s_{2}+s_{4}\right)
\right] ^{2}  \label{square} \\
&&+\frac{\varepsilon -1}{2}\left( \left( s_{1}+s_{3}\right) ^{2}+\left(
s_{2}+s_{4}\right) ^{2}\right) -2\varepsilon s^{2}  \nonumber
\end{eqnarray}
Since one knows how to combine representations of O(3), one easily obtains
the spectrum of $H_{square}$ for $N=3$. One finds a singlet for the ground
state with an energy of 
\begin{equation}
E_{square,3}=\left\{ 
\begin{array}{l}
\frac{\varepsilon }{2}-2\text{, }\varepsilon \leq 1 \\ 
-\frac{3\varepsilon }{2}\text{, }\varepsilon >1
\end{array}
\right\} 
\end{equation}
We assume in our calculation for $O(N)$, with arbitrary $N\geq 3$, that the
ground state continues to be a singlet, so $\left[ \left( s_{1}+s_{3}\right)
+\left( s_{2}+s_{4}\right) \right] ^{2}=0$ in eq(\ref{square}). It remains
to find the spectra of $\left( s_{1}+s_{3}\right) ^{2}$ and $\left(
s_{2}+s_{4}\right) ^{2}$. We now use standard Dirac Gamma Matrices instead
of real fermions in our argument and reduce the tensor product of two $O(N)$
spinor representations, say $\xi ^{\alpha }\phi ^{\beta }$, of $2^{\left[
N/2\right] }$ dimensions each, to calculate $\left( s_{1}+s_{3}\right) ^{2}$
and $\left( s_{2}+s_{4}\right) ^{2}$. The reduction of $\xi ^{\alpha }\phi
^{\beta }$ is achieved via antisymmetrised products of Dirac $O(N)$ gamma
matrices \cite{Georgi}: 
\begin{equation}
\xi ^{\alpha }\eta ^{\beta }\rightarrow \xi \eta \text{, }\xi \Gamma _{\mu
}\eta \text{, }\xi \Gamma _{\mu _{1}\mu _{2}}\eta \text{, ...,}\xi \Gamma
_{\mu _{1}...\mu _{N}}\eta   \label{reduction}
\end{equation}
where $\Gamma _{\mu _{1}...\mu _{n}}$ denotes a product of $n$ gamma
matrices that are antisymmetrised in their indices. The $O(N)$ spin of an
antisymmetric tensor $\Gamma _{\mu _{1}..\mu _{n}}$ is most easily found by
treating it as Grassmann numbers acted upon by $O(N)$ generators $%
s_{ij}=\Gamma _{i}\frac{d}{d\Gamma _{j}}-\Gamma _{j}\frac{d}{d\Gamma _{i}}$.
In this way one can find without too much difficulty that 
\begin{equation}
s^{2}\Gamma _{\mu _{1}..\mu _{n}}\equiv \frac{1}{2}%
\sum_{i,j=1..N}s_{ij}s_{ij}\Gamma _{\mu _{1}..\mu _{n}}=n(N-n)\Gamma _{\mu
_{1}..\mu _{n}}  \label{eigen}
\end{equation}
with a minimum eigenvalue of $s^{2}$at $n=[\frac{N}{2}]$ $\equiv \frac{N}{2}-%
\frac{1}{2}\delta _{N,odd}$ and with maxima at $n=0,N$. This results in the
following ground state energy of $H_{square}$ for arbitrary $N$:

\begin{equation}
E_{square ,N}=\left\{ 
\begin{array}{l}
-\frac{1}{4}N^{2}+\frac{\varepsilon }{4}N-\frac{1}{4}\left( \varepsilon
-1\right) \delta _{N,odd}\text{, }\varepsilon \leq 1 \\ 
-\frac{\varepsilon }{4}N^{2}+\frac{\varepsilon }{4}N\text{,}\varepsilon >1
\end{array}
\right\}  \label{groundstate}
\end{equation}
The ground state enery $E_{square,N}$ in eq(\ref{groundstate}) reduces
smoothly to its value at $N=3$, with a relative precision of order $\frac{1}{%
N^{2}}\sim 10\%$ for $N=3$, if one includes only the two leading terms, and
we conclude that the $O(N)$ extrapolation is satisfactory in this toy
example. By comparing with eq(\ref{dimer}) we notice that the leading $%
O(N^{2})$ term of $E_{square,N}$ can be interpreted in terms of the
formation of two dimers with coupling $\varepsilon $ or $1$, whichever is
larger. We shall see below that this is part of a more general pattern that
emerges at $N=\infty $.

\section*{Nature of the saddle point at N=$\infty $}

To use standard $N\rightarrow \infty $ techniques \cite{largeNpapers} we
rewrite the quartic interaction in terms of an auxiliary field \cite{Tsvelik}
\begin{eqnarray}
H &=&\sum_{<x,y>}J_{xy}(s_{x}s_{y}-\frac{N}{8})=\sum_{<x,y>}\frac{J_{xy}}{2}%
(\eta _{x}\eta _{y})^{2}  \nonumber \\
Z &=&\int D\eta e^{-\int_{0}^{\beta }dt\left( \frac{1}{2}\eta _{\mu
}\partial _{t}\eta _{\mu }+H\right) }=const(\beta )\int DBD\eta e^{-S} 
\nonumber \\
&=&const(\beta )\int DBe^{-\int_{0}^{\beta }dt(\frac{1}{4}\sum_{x,y}\frac{%
B_{xy}^{2}}{J_{xy}}-\frac{N}{2}\log (\det (\partial _{\tau }+iB))}  \nonumber
\\
S &=&\int_{0}^{\beta }dt\left( \frac{1}{4}\sum_{x,y}\frac{B_{xy}^{2}}{J_{xy}}%
+\frac{1}{2}\sum_{x,\mu }\eta _{x\mu }\partial _{t}\eta _{x\mu }+\frac{i}{2}%
\sum_{x,y}B_{xy}\eta _{x\mu }\eta _{y\mu }\right)   \label{aux}
\end{eqnarray}
With $J_{xy}$ scaling as $\frac{1}{N}$ the exponent is of order $N$ and the
integration over $B_{xy}$ is dominated by ''classical '' configurations of $%
B_{xy}$ plus fluctuations of $B_{xy}$ of order $O(1/\sqrt{N})$. To find the
optimal solution(s) of eq(\ref{aux}) at $T=0$ it is easiest to consider the
Hamiltonian that describes the saddlepoint of eq(\ref{aux}) and which is
given by 
\begin{equation}
H_{\infty }=\frac{1}{4}\sum_{x,y}\frac{B_{xy}^{2}}{J_{xy}}+\frac{1}{2}%
\sum_{x,y,\mu }iB_{xy}\eta _{x\mu }\eta _{y\mu }  \label{hinfinity}
\end{equation}
The optimal auxiliary fields $B_{xy}$ are those that give the lowest energy
and which minimise $<H_{\infty }>$. The matrix $iB_{x,y}$ is hermitean and
antisymmetric and its eigenvalues occur in complex conjugate pairs $(\xi
_{n},\lambda _{n}),(\xi _{n}^{*},-\lambda _{n})$, for $n=1,2..\frac{1}{2}%
\#(points)$. This property of $iB_{x,y}$ enables us to rewrite the kinetic
energy of the{\em \ real }fermions $\eta _{x\mu }$ in terms of {\em complex}
fermions with standard oscillator commutators: 
\begin{eqnarray}
\frac{1}{2}\sum_{x,y,\mu }iB_{xy}\eta _{x\mu }\eta _{y\mu } &=&\frac{1}{2}%
\sum_{n=1}^{\frac{1}{2}\#(points)}\sum_{\mu =1}^{N}\lambda _{n}\left(
A_{n,\mu }^{+}A_{n,\mu }-A_{n,\mu }A_{n,\mu }^{+}\right)   \label{oscillate}
\\
\text{with }A_{n,\mu } &=&\sum_{x=1}^{\#(points)}\xi _{nx}\eta _{x\mu }\text{%
, }  \nonumber \\
\left[ A_{m,\mu }^{+},A_{n,\nu }\right] _{+} &=&\delta _{mn}\delta _{\mu \nu
}\text{; }\left[ A_{m,\mu },A_{n,\nu }\right] _{+}=0\text{; }m,n=1..\frac{1}{%
2}\#(points)  \nonumber
\end{eqnarray}
We deduce from eq(\ref{oscillate}) that the zero point energy of the
fermions is given by 
\begin{equation}
<\frac{1}{2}\sum_{x,y,\mu }iB_{xy}\eta _{x\mu }\eta _{y\mu }>=-\frac{N}{4}%
\sum_{\text{all }\lambda }|\lambda |  \label{zeropoint}
\end{equation}
where one must sum over all $\#(points)$ eigenvalues. Combining eqs(\ref
{hinfinity},\ref{zeropoint}) we may rewrite the ground state energy at $%
N=\infty $ as 
\begin{equation}
E=\frac{1}{4}\sum_{x,y}\frac{B_{xy}^{2}}{J_{xy}}-\frac{N}{4}\sum_{\text{all }%
\lambda }|\lambda |  \label{start}
\end{equation}
In the case of frustrated magnetism we have, in general, distinct matrices $%
\frac{B}{\sqrt{J}}$ and $iB$ that do not commute. However, for a single
nonzero coupling $J_{xy}=J$ the energy associated with the auxiliary field
can be simplified and related to the spectrum of the matrix $iB$: 
\begin{eqnarray}
\sum_{x,y}\frac{B_{xy}^{2}}{J_{xy}} &=&\frac{1}{J}\sum_{x,y}B_{xy}^{2}=-%
\frac{1}{J}\sum_{x,y}B_{xy}B_{yx}\text{, }B_{yx}=-B_{xy}  \label{spectral} \\
&=&-\frac{1}{J}TrB^{2}=\frac{1}{J}Tr\left( iB\right) \left( iB\right) =\frac{%
1}{J}\sum_{\text{all }\lambda }\lambda ^{2}  \nonumber
\end{eqnarray}
Combining eqs(\ref{hinfinity},\ref{zeropoint},\ref{spectral}) we obtain a
lower bound on the ground state energy:

\begin{equation}
E=\sum_{\text{all }\lambda }\left( \frac{\lambda ^{2}}{4J}-\frac{|\lambda |N%
}{4}\right) =\frac{1}{4J}\sum_{\text{all }\lambda }\left[ (|\lambda |-\frac{%
JN}{2})^{2}-\frac{(JN)^{2}}{4}\right] \geq -\frac{JN^{2}}{16}\text{\#(points)%
}  \label{bound}
\end{equation}
We have used that the matrix $B_{x,y}$ has as many eigenvalues as there are
points on the lattice. We have thus obtained a lower bound for the energy of
an $O(\infty )$ antiferromagnet with a single coupling $J$.

For unfrustrated Heisenberg Hamiltonians on a lattice of points that can be
considered as a union of non overlapping pairs of points or ''dimers '' it
is easy to saturate this bound. For such Hamiltonians, we may choose a
configuration of $B_{xy}$ that is nonvanishing only on an arbitrary
collection of nonoverlapping dimers that cover the whole lattice. By
hypothesis, the couplings are the same on all these dimers. The matrix $%
B_{x,y}$ then decomposes into blocks, one for each dimer, and according to
eq(\ref{bound}) the energy can be rewritten as

\begin{eqnarray}
E &=&\frac{1}{4J}\sum_{\text{dimers}}\sum_{i=1,2}\left[ (|\lambda _{i}|-%
\frac{JN}{2})^{2}-\frac{(JN)^{2}}{4}\right] \stackrel{|\lambda |=\frac{JN}{2}%
}{\rightarrow }\frac{1}{4J}\sum_{\text{dimers}}\sum_{i=1,2}-\frac{(JN)^{2}}{4%
}  \nonumber \\
&=&-\frac{JN^{2}}{16}\cdot 2\cdot \text{\#(dimers)}=-\frac{JN^{2}}{16}\cdot 
\text{\#(points)}  \label{saturated}
\end{eqnarray}
Here we have adjusted the block of $B$ that corresponds to each dimer in
such a way that its eigenvalues are $\lambda =\pm \frac{JN}{2}$. We may also
return to equation (\ref{dimer}) to see more directly that the minimal
energy of a collection of nonoverlapping dimers (in their singlet state)
saturates the inequality (\ref{bound}): 
\begin{equation}
<\sum_{<x,y>}J(s_{x}s_{y})_{O(N)}>=-\frac{JN^{2}}{8}\cdot \text{\#(dimers)}%
+O(N)  \label{simple}
\end{equation}
which visibly saturates eq(\ref{saturated}). Returning to the toy model of
eq(\ref{square}) we now understand why its ground state energy corresponds,
to leading order in $N$, to that of two dimers, but our general arguments
apply only to $\varepsilon =1$ and $\varepsilon =0$ where there is a single
coupling. The toy model suggests the stronger statement that the ground
state on lattices coverable by dimers is one of singlets on the ''strongest
'' dimers.

To follow the hint of the toy model, we reconsider eq(\ref{start}) and
derive a lower bound for the ground state energy of a frustrated
antiferromagnets at $N=\infty $: 
\begin{equation}
E=\frac{1}{4}\sum_{x,y}\frac{B_{xy}^{2}}{J_{xy}}-\frac{N}{4}\sum_{\text{all }%
\lambda }|\lambda |\geq \frac{1}{4J_{\max }}\sum_{x,y}B_{xy}^{2}-\frac{N}{4}%
\sum_{\text{all }\lambda }|\lambda |  \label{major}
\end{equation}
Here we have used the positivity of the couplings $J_{xy}$. We may now copy
word for word the arguments that lead to the lower bound of eq(\ref{bound})
in the unfrustated case, but with $J$ replaced by $J_{\max }$. We find: 
\begin{equation}
E\geq -\frac{J_{\max }N^{2}}{16}\cdot \text{\#(points)}  \label{frustlow}
\end{equation}
In some cases this bound can be saturated and lowest energy saddle points
can be found. Consider, for example, the square lattice spin $1/2$
antiferromagnet, with nearest neighbor couplings $J_{1}$ and next nearest
neighbor couplings (across the diagonals) $J_{2}$. In this case we chose the
stronger coupling $J_{\max }=\max (J_{1},J_{2})$, and any dimer covering of
the lattice by the ''stronger bonds '' that correspond to $J_{\max }$. As
before, we saturate the lower bound with these configurations. So the
infinite degeneracy persists in this particular frustrated antiferromagnet.
The $N=\infty $ saddle point of an antiferromagnet on a Kagome lattice\cite
{Kagome} is also infinitely degenerate, because a Kagome lattice can be
covered by dimers in an infinite number of ways.
 However, there is only 
finite degeneracy in the $N=\infty $ saddle point of
the frustrated spin chains of \cite{Majumbdar}.

Although we have found an infinite number of degenerate saddle points in
certain $O(\infty )$ antiferromagnets we cannot be sure to have found {\em %
all} the saddle points that saturate the lower bound. In particular, the
saddle point may exhibit continuous degeneracies. A one parameter continuous
degeneracy is indeed present in the $O(\infty )$ saddle points of the
frustrated square of eq(\ref{square}), and an analogous degeneracy was also
noted in \cite{Read} for the case of $SU(\infty )$.

\section*{Conclusions and open problems}

We have found that a large class of $O(N)$ Heisenberg models have $N=\infty $
saddle points consisting of products of singlets and which are infinitely
degenerate in many cases.

Because there is a gap at large $N$ but no gap in the unfrustrated model at $%
N=3$ and $N=4$ there should be a critical value of $N$ that separates the
two regimes in the case of unfrustrated antiferromagnets, while for
frustrated antiferromagnets there is no need for such a phase transition.

Our results closely parallel those obtained by D.S. Rokhsar\cite{Rokhsar} on
the $N\rightarrow \infty $ limit of an $SU(N)$ extension of the Heisenberg
antiferromagnet. In the latter case, the effects of $1/N$ corrections are
understood\cite{Read}, while in the $O(N)$ case, the effect of these
corrections is still unknown. Also, it is still an open problem whether the
correlations can be usefully organised in powers of $N$ at $N=3$.

\section*{Acknowledgements}

We acknowledge encouragement by the members of CPTMB and thank S.Meshkov and
Y.Meurdesoif for their help and useful comments and Y. Chen for a critical
reading of the manuscript. F.T. acknowledges support of ENS de Lyon. D.F.
acknowledges support by the ESPRIT / HCM / PECO - Workshop at the Institute
for Scientific Interchange Foundation, Torino, Italy, and discussions with
N.Schopohl.

\end{document}